\documentclass[sigconf]{acmart}
\usepackage{hyperref}
\usepackage[hyphenbreaks]{breakurl}

\AtBeginDocument{%
  \providecommand\BibTeX{{%
    \normalfont B\kern-0.5em{\scshape i\kern-0.25em b}\kern-0.8em\TeX}}}

\copyrightyear{2023}
\acmYear{2023}
\setcopyright{rightsretained}
\acmConference[CHIWORK 2023]{Annual Symposium on Human-Computer Interaction for Work 2023}{June 13--16, 2023}{Oldenburg, Germany}
\acmBooktitle{Annual Symposium on Human-Computer Interaction for Work 2023 (CHIWORK 2023), June 13--16, 2023, Oldenburg, Germany}\acmDOI{10.1145/3596671.3598576}
\acmISBN{979-8-4007-0807-7/23/06}%
\author{Jane Hsieh}
\email{jhsieh2@cs.cmu.edu}
\affiliation{%
  \institution{Carnegie Mellon University}
  \country{USA}
}

\author{Oluwatobi Adisa}
\email{oadisa@andrew.cmu.edu}
\affiliation{%
  \institution{Carnegie Mellon University}
  \country{USA}
}

\author{Sachi Bafna}
\email{sachib@andrew.cmu.edu }
\affiliation{%
  \institution{Carnegie Mellon University}
  \country{USA}
}

\author{Haiyi Zhu}
\email{haiyiz@cs.cmu.edu}
\affiliation{%
  \institution{Carnegie Mellon University}
  \country{USA}
}

\begin{document}

\title{Designing Individualized Policy and Technology Interventions to Improve Gig Work Conditions}

\begin{abstract}
The gig economy is characterized by short-term contract work completed by independent workers who are paid to perform ``gigs'', and who have control over when, whether and how they conduct work. Gig economy platforms (e.g., Uber, Lyft, Instacart) offer workers increased job opportunities, lower barriers to entry, and improved flexibility. However, growing evidence suggests that worker well-being and gig work conditions have become significant societal issues. In designing public-facing policies and technologies for improving gig work conditions, inherent tradeoffs exist between offering individual flexibility and when attempting to meet all community needs. In platform-based gig work, contractors pursue the flexibility of short-term tasks, but policymakers resist segmenting the population when designing policies to support their work. As platforms offer an ever-increasing variety of services, we argue that policymakers and platform designers must provide more targeted and personalized policies, benefits, and protections for platform-based workers, so that they can lead more successful and sustainable gig work careers. We present in this paper relevant legal and scholarly evidence from the United States to support this position, and make recommendations for future innovations in policy and technology. 
\end{abstract}





\maketitle

\section{Introduction}
Gig work is often characterized by short-term and digitally mediated tasks or projects performed by independent contractors, who usually hold control over where the work is done \cite{Watson2021-kr}. Recent accounts show that gig workers comprise a sizable and rapidly growing segment of the nonstandard workforce, revolutionizing the informal workforce by providing a cost-effective earning opportunity for workers with low skills or education, enabling small businesses to scale quickly, and offering greater mobility to disadvantaged citizens \cite{Wood2018-lq,Branstetter2020-gw,Blaising2021-ue}. By 2021, more than one-third of the US working population had participated in gig work \cite{noauthor_undated-xj}, and more than 3\% of adults engaged with it as their primary occupation \cite{Anderson2021-pa}. But alongside the proliferation of platforms, scholars are expressing concerns about how algorithmic management affects the well-being of gig workers \cite{You2021-wb,Zhang2022-ji,Lee2021-eb,Cram2020-cf,Cram2022-pf,Wu2022-rk,Berger2019-yt,Parry2019-gj}.

Previous studies report the limited social \cite{Yao2021-eb,Kuhn2019-fi,Wood2019-jp}, technological \cite{Li2022-hd,Jarrahi2020-zz}, and regulatory \cite{Dubal2019-qi,Hardy2022-xd,Graham2017-zw,Tan2021-li,Stewart2017-bs} support needed to contend with the adverse conditions of gig work, which include intense competition \cite{Bates2021-sv,Kost2020-gb}, low wages \cite{Friedman2014-cz,Kalleberg_undated-sb,Wood2018-lq}, job precarity \cite{Ashford2018-dw,Sutherland2020-wk}, and physical hazards \cite{Bajwa2018-gy,Howard2017-wd}. These inadequacies arise from several unique characteristics of workers, such as information asymmetries \cite{Jarrahi2019-if,Jarrahi2020-zz,Zhang2022-ji,Kinder2019-pr}, a commodification of labor \cite{Stefano_undated-dw,Wood2019-jp}, and a lack of benefits and protections \cite{Muntaner2018-dn,Luckett_undated-qr}. Underlying these problems is the ambiguous legal classification of gig workers-their status as independent contractors provides flexibility in the time and location of work, but at the expense of typical employee rights to unionization and collective bargaining or benefits such as health care and paid time off \cite{Bales2016-df}.

In light of these concerns, many have turned to the redesign of regulatory policies [39] and platforms \cite{Arnoldi2021-xq,Jabagi2019-dt}. However, many existing proposals suggest broader advancements that uniformly benefit the entire gig workforce without considering individual workers' unique constraints and priorities \cite{Hsieh2023-cu}. Such general, all-encompassing solutions overlook the diversity of gigs and the population of workers who complete them. Many different categories of gig work exist \cite{Duggan2020-qh}, and along with them unique occupational hazards and worker needs  \cite{Bajwa2018-gy}. While gig workers may bear the same risks as others doing similar work outside of platforms (e.g. taxi drivers), they do not share the benefits and protections typically afforded to employees. 

In this position paper, we argue that instead of universal benefits and solutions, policy and platform designers need to consider more targeted and personalized policies and features to support the unique needs of individual workers, who undertake a wide variety of tasks types and occupy diverse and intersectional backgrounds. For example, policymakers of the U.S. can amend existing codes and introduce new legislation to provide workers with collective bargaining power, protections against discrimination and retaliation, as well as specialized bills for addressing particular working needs of different types of gig workers. On the platforms' end, engineers and designers can implement features and services that target and accommodate individual needs, increase worker agency, and in general improve worker well-being and welfare. Outside of platform initiatives, technological advancements such as automation can assist workers with tax filing, financial tracking, contract writing as well as promoting individual well-being.

\section{Background}
\subsection{Diversity of Gig Services and Participants}
Gig work is multifaceted and encompasses many service industries, ranging from physical labor such as construction work to digitally deliverable services such as logo design or software development. While some primary attributes (e.g., placement platform or spatial/temporal flexibility) unify and define all forms of gig work, further classifications can identify multiple variants and categories. Some studies broadly divide gig work into its physical and remote counterparts \cite{Fauzi2022-vd,Hsieh2022-bl}. For example, De Stefano distinguishes between crowdwork – platforms such as Amazon Mechanical Turk or CrowdFlower that mediate the remote execution of microtasks – and app work – intermediaries such as Uber or TaskRabbit that connect workers to tasks performed locally, including transportation, cleaning, and various other errands such as food delivery \cite{Stefano_undated-dw}. Duggan also introduced capital platform work as a third variant, characterizing the work of online sellers who use digital platforms such as Etsy and Airbnb to share individually-owned capital with consumers \cite{Duggan2020-qh,Maffie2020-ap}. In addition to task-based classifications, a literature review by Watson et al. also profiled different groups of gig workers: the Gig Service Provider (e.g., Uber, AirBnb, TaskRabbit app workers), the gig goods provider (e.g., online sellers such as Etsy or RedBubble), the gig data provider (e.g. crowdworkers like AMT or Google Surveys), the Agency Gig Worker (where contractors are assigned work through an intermediary agency), and the Traditional Gig Worker, which includes roles such as substitute teachers, comedians, babysitters, photographers, and musicians \cite{Watson2021-kr}.

Each of the above gig work variants require a distinct set of protections and resources. In addition to work-related specializations, workers also diverge from one another in terms of demographics. Recent surveys show that gig workers tend to be younger, male, Hispanic/Black, more likely to be educated, and live in urban areas compared to traditional workers \cite{noauthor_undated-tq,Gelles-Watnick2021-fz}. In contrast, an older report from BLS in 2005 showcased an older and more white gig workforce \cite{Howard2017-wd}. Within each category of gig work, however, there are more subtle differences: independent contractors tend to be older and whiter, workers on online platforms tend to be male, and women are more likely to engage in capital platform work \cite{Davidson2022-le}. This diversity of demographic and occupational characteristics (and the associated intersection between groups) necessitates the development of targeted policies and platform features that address the specific work needs of each group. In the following sections we outline the current state of challenges faced by gig workers as well as existing forms of regulatory support and budding policy initiatives to address such issues.

\subsection{Challenges and Inequalities in Gig Work}
Currently, gig workers of the United States are exposed to various financial, safety, and health risks while lacking access to many forms of necessary social, technical, and legal support. Previous work suggests that platforms are unwilling to implement programs and features to improve working conditions due to the high costs involved, and that policymakers tend to favor one-size-fit-all solutions that apply to all gig workers to ensure inclusion and avoid segregation \cite{Hsieh2023-cu}. However, not only are such homogeneous solutions hard to devise, they will not meet the individual needs of gig workers, and platforms cannot be expected to provide the benefits required by each individual worker. Below, we outline some key concerns and priorities for specific subgroups of workers, and subsequently propose ways of expanding existing U.S. policies and platform support to address such shortcomings.

\subsubsection{Occupational Hazards of (Physical) App Workers}
App workers performing physical work (e.g., ride-sharing, delivery services, contractual construction work) need safety precautions and safeguards, access to bathrooms, and flexibility in determining where they work (by setting a maximum radius of travel etc.). A survey of 4,000 gig workers conducted by UCLA found that 37\% of delivery drivers have suffered an accident while on the job \cite{noauthor_undated-is}. To make matters worse, drivers are disincentivized from taking protective measures, such as using dashcams, as passenger discomfort with surveillance can lead to poor ratings, which are important inputs to platforms' rating systems \cite{Almoqbel2019-in}. According to NIOSH, delivery drivers face a higher risk for work-related motor vehicle accidents compared to workers in other occupations \cite{noauthor_1998-ts}. Many cases of violence likely go undetected, but the Markup has identified 361 ride-hail and delivery drivers as victims of carjackings or attempted carjackings over just the last five years alone \cite{Kerr_undated-zw}. Cleaners, caregivers, and TaskRabbit workers also face the dangers of entering strangers' homes to offer their services \cite{Bajwa2018-gy}. These workplace hazards are exacerbated by the lack of training and on-site supervision expected in traditional work arrangements.

In addition to physical safety risks, psychological and physical work demands placed on app workers further expose them to health risks \cite{Keith2020-zb}. Studies of mortality and psychological morbidity studies have shown that workers in nonstandard, gig work arrangements are at higher risk of physical and mental injuries than workers in standard industrial work environments \cite{Howard2017-wd}. Empirical studies have also found that job insecurity (a typical characteristic to contingent workers ) has a negative impact on health and well-being \cite{De_witte2016-wz}.

\subsubsection{Conditions of Online Gig Work}
Unlike location-dependent app workers, online-based workers like freelancers are more likely to deal with uncertainty in payments and opportunities \cite{Kitamura2021-gh,Blaising2021-ue} and invasive monitoring \cite{Popiel2017-wu}. \textbf{Freelancers} experience platformic management when their performance evaluations are documented in ranking systems (which depend on client reviews), as well as extensive oversight when their keystrokes and active time are recorded \cite{Vega2022-qv}. In addition, international clients subject freelancers to long and unusual work hours \cite{Shevchuk2021-qh,Shevchuk2019-vb}, which increases emotional exhaustion, leads to a blended work-life balance, and undermines life satisfaction. Finally, freelancers are responsible for their own reputational management, which can entail extensive time spent on building profiles and maintaining positive relationships with clients \cite{Kitamura2021-gh,Hsieh2022-bl}.

\textbf{Crowdworkers} face similar challenges of self-management and long working hours. However, they must additionally endure the challenges of unfair pay \cite{Barbosa2019-kp} and wage theft, as their payment relies on the approval of requesters and a relatively large supply of laborers makes their services fungible and easily replaceable \cite{Irani2013-pf}. Such imbalance of labor supply and demand also creates meaningless, menial tasks as well as low pay and recognition \cite{Kittur2013-jd,Durward2020-vh}.

\textbf{Online goods providers} are more vulnerable to hikes in transaction fees, or competition from large corporate companies \cite{West_undated-no}, although they also face the challenges of low pay, algorithmic management, and the pressure of reputation upkeep through rating systems \cite{Benson2020-it}. For peer-to-peer sharing platforms such as AirBnb, reviews constitute yet another performance metric that workers must work to maintain \cite{Lawani2019-ry}. Finally, due to invisible platform policies, online sellers often have to negotiate to defend that their products amount to ``handmade'' commodities and familiarize themselves with intellectual property laws to defend against infringement \cite{Razaq2022-lq}. 

\subsubsection{Reinscriptions of Inequality in Gig Work}
\textbf{Gender} pay gaps are reported in both crowd work and freelancing \cite{Dunn2021-hq,Foong2021-ho,Dubey2017-fy,Litman_undated-kh}, requiring women to work more hours on platforms to make up for the differences in pay \cite{Barzilay2016-yl}. In addition to unfair remuneration, occupational gender stereotypes are perpetuated in various gig work sectors around the world \cite{Galperin2021-eh,Wood2018-lq}. However, due to caregiving and other domestic responsibilities, women are less able to fulfill the long working hours demanded of freelancers \cite{Adams-Prassl2017-co}. To top it off, harassing behaviors such as verbal abuse, stalking, or bullying are more likely to put women at risk \cite{Rosenblat2017-bm}. The lack of public or platform-enforced anti-harassment policies have led women workers to resort to ``brushing it off'' when harassment does occur \cite{Ma2022-ba}, or to use usernames that don't reveal their gender \cite{Kasliwal2020-zg}.

Despite the disproportionate participation of Black and Hispanic populations in gig work, occupational segregation and \textbf{racial discrimination} remain prevalent \cite{England_undated-gk}. On AirBnb, profile pictures have resulted in Black hosts charging 12\% less than non-Black hosts \cite{edelman2014digital}. In an experimental study of hypothetical hiring decisions, Black candidates were 16\% less likely to be hired \cite{Leung2020-rk}. A 2021 survey found that White workers were less likely than their white counterparts to earn from multiple types of gig jobs (48\% vs. 30\%), to feel unsafe while completing jobs (41\% vs. 28\%), and to receive unwanted sexual advances (24\% vs. 13\%) \cite{Gelles-Watnick2021-fz}. There is also evidence of disparities between goods providers on Craigslist and eBay, where a White person's hand in product photos helped garner higher prices than a Black one \cite{Rosenblat2017-bm}.

\textbf{Socioeconomic factors} may underlie many existing inequalities in the gig economy \cite{Shaw2022-ai}. Compared to low socioeconomic status (SES) areas and the suburbs, services such as UberX and TaskRabbit were found to be significantly more effective in dense high-SES areas \cite{Thebault-Spieker2017-bv}. AirBnB workers tend to have higher education, higher income, and strong ties in the labor market, forming a barrier to entry for individuals of lower socioeconomic status \cite{Ilsoe_undated-mr}. These findings are noteworthy because studies point to the potentially detrimental impacts of SES on late-life poor health outcomes, such as aging and mortality \cite{Freni-Sterrantino2021-fq,Pathak2022-ah}.

A discussion of inequalities would be remiss to not consider the dynamics of intersectionality. Workers perceived as women on TaskRabbit (especially white women) received 10\% fewer reviews, and Black (men) received significantly lower ratings; Black men on Fiverr also received ~32\% fewer reviews \cite{Hannak2017-xn}. While past studies have substantiated claims of racial, gender and socioeconomic biases, we lack understanding (and therefore encourage future investigations) around gig works' impacts on other vulnerable groups, such as the disabled and older populations, as well as intersecting inequalities (e.g. the gendered experiences of workers in low- and middle-income countries) \cite{Barakat2022-tz,Hunt2019-qb}.

\subsection{Existing Initiatives and Preliminary Policies}
Many of the work-induced challenges described above have fueled legislative concern in the US, where the state of California leads the nation's disagreement around labor laws. In January of 2020, the Assembly Bill 5 (AB5) amended the state's labor laws to expand the definition of an employee so as to reduce the chances of employers misclassifying regular workers as independent contractors. The bill was subsequently mandated by courts and extended labor protections like paid leave to an estimated one million people \cite{noauthor_2021-px}. However, in the November 2020 state election, a ballot initiative (which cost platforms more than 200 million in campaign funds \cite{Stecker2020-xk}) was passed to exempt app-based transportation companies from AB5 \cite{Hussain2022-lr}. In August 2021, the initiative was declared unconstitutional and unenforceable by a county court judge \cite{Gedye2023-ru, unconstitutional}, but proponents of Prop 22 subsequently appealed the ruling in December 2022; they are expected to receive a decision from the California Supreme Court \cite{Yakal2023-ka}.

Other states also strive to bring more benefits to gig workers. In March 2022, Washington state governor Jay Inslee signed the Engrossed Substitute House Bill 2076 into state law (in effect by January 2023), which guarantees minimum trip payments, workers' compensation, paid sick leave (one hour earned for 40 worked hours), as well as a resource center to educate workers on received benefits \cite{Fletcher2022-qr}. A pair of proposed ballot initiatives would guarantee Massachusetts drivers benefits such as minimum wage, per-mile expense reimbursements, a health care stipend, paid leave, workers' compensation, protection against discrimination, as well as a right to appeal terminations. At the federal level, a bipartisan group of legislators from the US House of Representative introduced a federal bill in July 2022 that would make it compulsory for platforms to provide a written summary of worker benefits, allowing workers to reject assignments and conduct multi-platform work, affording them rights to privacy, safety and leave as described in the Family and Medical Leave Act, as well as protections against client discrimination, retaliation and harassment \cite{Shepherd2022-ed}.

Besides initiatives to guarantee worker benefits, a plethora of classification tests are getting developed and adopted to assess the appropriate classification of workers. Since 2019, California and nine other states have adopted (or are considering) the employee-friendly ABC test to avoid misclassification of workers as independent contractors \cite{Iacurci2022-po}. More recently, President Biden also proposed a national rule to test whether a gig worker could be considered an employee based on factors the amount of control workers have over how they conduct work as well as the opportunities to increase earnings by offering new services \cite{Scheiber2022-oo}. 

\subsection{Gaps between Worker Needs and Existing Policies}
The existing bills and proposals take many large-scale issues (e.g. workers' compensation and paid leave) into consideration, but they do not make provisions for the needs of specialized communities, excepting rideshare drivers. Many subgroups of workers await targeted policies to assist with particular dimensions and issues of their work – working mothers need paid maternal leave while disabled and marginalized workers require public accommodations for meeting various health and safety needs. For instance, our past work eliciting the perspectives of multiple stakeholders found platforms to resist the implementation of worker benefits, advocating instead for worker-initiated collective actions, and that existing public infrastructure failed to provide for basic working needs of gig workers \cite{Hsieh2023-cu}. Platform reluctance to provide benefits stemmed from the fear of being imposed a legal employment relationship with the worker (as consistent with prior work \cite{Harris_undated-xb}). On the other hand, these workshops also revealed how regulators preferred all-inclusive solutions to special accommodations (e.g. universal healthcare or universal basic income) to minimize the risk of excluding (potentially vulnerable) segments of the population, and avoided personalized solutions as they pose potential threats to worker privacy \cite{Hsieh2023-cu}. However, worker participants of the workshops voiced desires for customized solutions to meet individual needs, the agency to leverage multiple platforms or conduct their own financial planning, as well as to avoid classification as employees \cite{Hsieh2023-cu}. Finally, the investigation of an ``indie'' food delivery system by Dalal et. al. uncovered how platforms that prioritize local contexts over transnational scales offers unique affordances and possibilities for workers \cite{dalal2023understanding}.

While an argument might be made for an omnibus bill that includes all benefits and protections to address needs of all gig workers \cite{Harris_undated-xb}, such a tendency toward all-encompassing policies and benefits have the downside of being very broad, leading to a high cost associated with their implementation and major labor revisions. Cost is already a major reason for platforms' inhibition against implementation of benefits \cite{Hsieh2023-cu}, and also contributes to the lengthy process of policy implementation \cite{Hudson2019-un}. Presenting specialized policies that target specific issues can reduce the legislative burden in terms of which committees and jurisdictions to involve, potentially allowing for the earlier and faster presentation of highly-prioritized benefits. 

\section{Envisioned Advancements}
Prior work has highlighted the need for a third category of workers to lift legal ambiguity, improve working conditions, and increase market efficiencies \cite{Harris_undated-xb}. Currently, workers can choose among the binary categories of the formal employee or the independent contractor. But attempting to force the newer and more informal working arrangements of the gig economy into such pre-existing categories limits real and potential economic benefits of short-term, contractual workers, and recent findings show that multiple involved stakeholder groups advocate for the establishment of a new legal class of workers \cite{Hsieh2023-cu}. Harris and Krueger terms this third class the ``independent workers'', and argues for their various social and economic benefits \cite{Harris_undated-xb}. In the following, we examine existing models of relevant (and sometimes specific) policies at the local and federal level in the US and propose ways in which they can be extended, improved or adapted to benefit other groups of gig workers as well. 

\subsection{Policy Innovations}
\subsubsection{Power to Collectively Bargain}
Currently, gig worker communities are fragmented and blocked from socializing and forming a collective identity due to a variety of factors including platformic design, legal constraints as well as fears of platform retaliation. The main legal impediment for collective bargaining among gig workers is federal antitrust law, which states that ``Every contract, combination \dots in restraint of trade or commerce \dots is declared to be illegal.'', hence ``Every person who shall make any contract or engage in any combination or conspiracy hereby declared to be illegal'' (15 U.S Code § 1). These laws were codified in an attempt to outlaw monopolistic practices, so as to keep a free competitive market with low prices and high quality. Since gig workers are largely classified as independent contractors rather than employees, antitrust laws prohibit their efforts to collectively organize and bargain – any of their attempts at collective action can be treated as illegal cartelistic behaviors by courts. However, legal employees hold a ``labor exemption'' from antitrust liabilities since ``The labor of a human being is not a commodity or article of commerce'', and so ``Nothing contained in the antitrust laws shall be construed to forbid the existence and operation of labor \dots organizations'' (15 U.S. Code § 17). Employee efforts to collective action and unionizing are further protected by the National Labor Relations Act, which states that ``Employees shall have the right to self-organization, to form, join, or assist labor organizations, to bargain collectively through representatives of their own choosing, and to engage in other concerted activities for the purpose of collective bargaining or other mutual aid or protection'' (29 U.S. Code § 157). 

We argue that gig workers, much like formal employees, should also be granted exemption from antitrust laws since many of the necessary benefits and protections they require for work (outlined in above sections) can be easily negotiated once workers gain power to collectively bargain. Prior work revealed that regulators strongly supported workers to collectively bargain for their needs \cite{Hsieh2023-cu}, and such arrangements would empower workers to protest unfair working conditions, as well as benefit societal welfare at large by facilitating more efficient and rapid allocation of market resources. Furthermore, workers would more rapidly gain access to benefits (as compared to the time-consuming process of implementing specific legislation), and different groups of workers can have the flexibility and agency to prioritize benefits according to specific working needs. For instance, individuals completing gigs in food delivery and ridesharing are more likely to bargain for workers' compensations and bathroom access whereas freelancers and crowdworkers might negotiate for higher wages. Such benefits negotiated from worker-initiated action are more flexible and efficient than policy amendments as they can be more quickly negotiated and updated. The exemption can be applied to gig workers a few different ways: a new, third federal category of workers can be created to adapt labor laws to the changing nature of work, gig workers can be reclassified as employees, or workers can be directly granted exemption on a case-by-case basis.

\subsubsection{Specialized Policies for the Food Delivery Industry}
In September 2021, the New York City Council took leading steps in improving the welfare of food delivery workers by passing six bills that provide them benefits and protections through actions from the service platforms as well as a city agency \cite{noauthor_2021-px}. In terms of actions from the city, Bill Int No. 2294-2021 requires ``the Department of Consumer and Worker Protection to study the working conditions of third party food delivery workers'', and ``based on the results of the study \dots no later than January 1, 2023, the department shall by rule establish a method for determining the minimum payments that must be made''. Regarding payment, Int. No. 2296 mandates that platforms ``shall not charge or impose any fee on a food delivery worker for the use of any form of payment'' (thereby removing additional fees) and that that worker will be paid ``for work performed no less frequently than once a week''. With respect to tips, Bill Int. No 1846-2020 prohibits platforms from ``solicit[ing] a gratuity for a food delivery worker \dots unless such third-party food delivery service discloses, in plain language and in a conspicuous manner \dots amount of each gratuity that is distributed \dots whether such gratuities are distributed immediately \dots and whether such gratuities are distributed in cash''. Furthermore, the bill states that ``For each transaction \dots [the worker] shall be notified of how much the customer paid as gratuity'', and overall platforms must disclose to workers ``the aggregate amount of compensation \dots gratuities earned'', essentially requiring transparency for tips. To increase worker agency, Bill Int. No. 2289 states that workers will be provided ``the ability to specify: the maximum distance per trip \dots`` as well as parameters that allow workers to ``not accept trips that require travel over any bridge or \dots tunnel''. Finally, bills 2298 and 2288 and equip workers with physical accommodations: Int. No. 2298 mandates that ``a toilet facility is available for the use of food delivery workers lawfully'' on premises of food service establishments (i.e. restaurants), provisioning workers with access to necessary bathroom facilities, and Int. No. 2288 makes ``available insulated bags to any delivery worker who has completed at least six deliveries for the company''.

This exemplary model addresses specific needs of gig workers in New York City and we believe that it could be applied toward many other locations and sectors for a similar effect. For instance, grocery deliverers such as Instacart shoppers complete adjacent work, and can also benefit from the transparency in tips, set maximum distance per trip, access to bathrooms, as well as provisions of insulated bags. The requirement to pay workers at least once a week can be broadly applied to many worker types, including crowdworkers, online sellers and freelancers. Finally, workers in other critical industries of the gig economy (e.g. ridesharing or healthcare) can similarly benefit from a specific and targeted set of public policies \cite{Sumagaysay2022-ya}.

\subsubsection{Anti-discrimination Protections}
Many suggest that labor platforms provide income opportunities and  increased mobility for  disadvantaged workers who are otherwise incapable of engaging in full-time jobs (e.g. mothers, students, or individuals from marginalized and undereducated communities), offering them income through low-entry gigs as well as low-cost services \cite{Choudary2018-qn,Branstetter2020-gw}. But as members of marginalized communities are becoming increasingly involved in platformic and precarious work \cite{Dillahunt2021-ky}, they are also exposed to more risks that are inherent to short-term precarious work. Thus, policy amendments are necessary to adapt to the changing nature of work so that all gig work individuals have access to equal opportunities and necessary protections. 

As a basis for comparison, workers of the formal economy are well protected from discriminatory employment practices by federal statutory protections enforced by the Equal Employment Opportunity Commission. For instance, Title VII of the 1964 Civil Rights Act forbids employers and employment agencies from refusing ``to hire or to discharge \dots or \dots discriminate against any individual with respect to \dots compensation, terms, conditions, or privileges of employment, because of such individual's race, color, religion, sex, or national origin'' or ``to limit, segregate, or classify \dots employees or applicants \dots in any way \dots deprive any individual of employment opportunities \dots because of such individual's race, color, religion, sex, or national origin''. Beyond restricting discriminatory hiring and firing practices, sections of Title VII also cover actions related to promotions, compensation, training decisions, job shift assignments, merit systems, or disparate impact cases, among others \cite{noauthor_2019-kq}.

On the other hand, gig workers are only granted limited protections due to their current status as independent contractors. In particular, Section 1981 of the 1866 Civil Rights Act states that ``All persons \dots shall have the same right \dots to make and enforce contracts, to sue, be parties, give evidence, and to the full and equal benefit of all laws and proceedings for the security of persons and property as is enjoyed by white citizens'' \footnote{Where ``make and enforce contracts'' includes the making, performance, modification, and termination of contracts, and the enjoyment of all benefits, privileges, terms, and conditions of the contractual relationship.}, thereby prohibiting race-based discrimination when making, performing under, and terminating contracts. A subsequent amendment by Congress clarified that Section 1981 applies to private (as well as state) instances of racial discrimination ``in all forms of contracting, no matter how minor or personal'' \cite{Leong2016-eb}. While important, this section is severely limiting in protecting marginalized populations of gig workers as it only allows individuals to bring federal claims if platforms discriminate on the basis of race, but not other characteristics such as gender, age, disability, religion or ethnicity. In order for such protections to reach the various marginalized populations, we propose gig workers should also be protected under federal employment laws against discrimination, and similar to the right of collective bargaining, this can be achieved through either worker reclassification, a new class of workers, or case-by-case applications.

\subsubsection{Expansion of Anti-retaliation Protections}
Wage theft refers to situations where clients fail to pay for work that has been completed or pay less than the agreed-upon amount. This can have serious consequences for workers, including financial strain, loss of income, and reputation damage. For employees, the Fair Labor Standards Act (FLSA) ensures rights such as minimum wage and is enforced by the U.S. Department of Labor \cite{flsa}. The FLSA also provides employees anti-retaliation protections, so that employers cannot take adverse actions (e.g. firing) against employees who report misconduct or violations of labor laws. 

Unfortunately for gig workers, by 2016 nearly two-thirds of platforms included a forced arbitration agreement, which requires workers to submit disputes or reports of violations to arbitrators instead of to court, which means that workers cannot bring their own lawsuits to recover unpaid wages or other damages \cite{noauthor_2021-ms}. Furthermore, almost all of these agreements included a class action waiver, which bans workers from bringing their claims as a group in arbitration, even if the claims are borne of the same unlawful workplace practices. Thus, we suggest the creation of a private right of action for gig workers, so that individuals can pursue legal remedies if their rights under the FLSA are violated. Online gig workers (e.g. crowdworkers and freelancers) are especially vulnerable to wage theft or delayed payments due to the digitally mediated nature of their work, and thus would benefit from an expansion of such anti-retaliation efforts to protect their rights to speak up and report unfair treatments.

\subsection{Technological Gaps}
While implementing public or platform policies can lead to significant improvements in working conditions, we also recommend some technological advances tailored toward individual workers that can be developed alongside policy. These proposed technological improvements can either be integrated into platforms (as in the case of in-app customizations) or exist as external resources (financial planners, automated tax-filing). Unlike the quality-of-life improvements that policy amendments/additions may bring, the technical innovations, features and extensions might only result in incremental changes. Nonetheless, personalization is most feasibly achieved at the individual level, and technological approach centering end-users may constitute the most suitable and practical way of addressing users' diverse preferences \cite{Li2022-hd}.

\subsubsection{Platform-initiated Features}
Many of the above measures are implementable by platforms even if not required by policy mandates. For instance, increased transparency in gratuity and frequency in pay disbursements can benefit many workers of the on-demand and online sectors. In addition to fair practices in remuneration, platforms might consider systems that tailor toward worker preferences and schedules. Customizations may be implemented for workers to express their desired schedules and tasks preferences – the choice of opting out of deliveries that requires crossing tunnels/bridges constitutes one such personalization. On top of tailored options, in-app earnings projections can also assist individual workers in conducting personal financial planning. As gig work is making earning opportunities available to otherwise unemployable individuals (e.g. mothers, students, disabled persons), it is of increasing importance to elicit and subsequently accommodate each workers' unique needs, priorities, constraints, and context. 

\subsubsection{Developments Designed for Individual Well-being}
After gaining an understanding of workers' needs and objectives, platforms (or outside efforts) can incorporate various features to help improve individual well-being. For example, past work by You et. al. has developed a ``social sensing'' probe that shares drivers' personal health data with their significant others so as to increase awareness and establish accountability for maintaining well-being \cite{You2021-wb}. Zhang et. al. has similarly investigated how data probes (interactive data visuals) can surface individual workers' well-being and positionalities, which affect working strategies \cite{zhang2023stakeholder}.
Such developments, combined with nudges and reminders, can help workers collectively resist irregular schedules as well as long working hours. 

While some protective measures against late or non-payment are already in place (e.g. escrow), dynamic or smart contracts offer another way to protect worker wages. Smart contracts that self-execute and auto-enforce based on predefined conditions can help with the lack of transparency and wage theft, allowing freelancers to automatically receive payments that trigger when the conditions are met, reducing payment conflicts. Time and effort required to manage contracts will also decrease, helping mitigate long work hours. Finally, there is an opportunity gap for automation since independent contractors are required to file quarterly tax payments per year -- AI-powered assistance with financial tracking and tax codes could help alleviate this burden, freeing up time and effort.

\section{Concluding Discussion}
In this position paper we discuss ways that policy and technologies in the United States can target and improve the well-being of individual workers of the gig economy. We argue that specialized policies and personalized technology are efficient ways of addressing worker needs and promoting well-being. This work contributes to the CHIWORK community by 1.) presenting a summary of existing literature on diverse gig worker types, needs and populations 2.) analyzing existing legislative efforts from both legal and social perspectives and 3.) making recommendations for ways of extending the current policy and technological developments.

\subsection{Positionality, Limitations and Future Work}
We acknowledge the existence of potential drawbacks to personalized features or specialized policies that we did not consider or discuss. Although the authors of this paper vary in educational backgrounds (our training lies within Human Computer Interaction, Design, Information Systems, Software Engineering and Computer Science), gender as well as race, we all currently live within the US and only one author has resided outside the country within the past five years. As a result, our analysis centers specifically on the U.S. context, yet many platforms operate in a variety of countries across multiple continents. This regional bias may have caused us to overlook existing policy solutions in other regions, or more seriously, we may miss critical issues that are only present in non-US regions, such as the global south. We invite future lines of inquiry to evaluate the degree to which existing policy and technology serves the individualized needs of gig workers in other regions -- the Europe-based investigation on workers' data access rights by Stein and Calacci is a good example \cite{stein2022workers}. Secondly, we asserted that collective bargaining would more efficiently bring worker benefits, but this line of argument assumes the existence of union members who are willing to spearhead collective initiatives –- it's possible that workers are still reluctant to damage their relationships with platforms even after receiving exemption from antitrust laws. The legal classification of gig workers is another contentious issue that our proposed policy amendments does not address, but it is in fact a critical issue that merits further discussion and consideration. We hope that this paper furthers the community's understanding of potentials and tensions surrounding gig worker well-being and helps generate ideas for the key issues raised on working conditions. Future lines of work and dialogues are welcomed to extend or push back on this perspective towards individualized ways of promoting gig worker well-being.

\begin{acks}
This material is based upon work supported by the National Science Foundation (NSF), in part under Award 1952085, and in part by the Graduate Research Fellowship Program under Grant No. DGE2140739. Any opinions, findings, and conclusions or recommendations expressed in this material are those of the author(s) and do not necessarily reflect the views of the National Science Foundation.
We are thankful to Professor Lee Branstetter for providing invaluable advise on how to approach the research related to policy-related innovations.
\end{acks}

\bibliographystyle{ACM-Reference-Format}
\bibliography{references}

\end{document}